\documentclass{aa}
\usepackage[varg]{txfonts}
\usepackage{longtable}
\usepackage{lscape}
\usepackage{graphicx}
\newcommand{\Ch}    {\mbox{$M_{\rm Ch}$}}

\newcommand{\kmprs}  {\mbox{\rm km\,s$^{-1}$}}
\newcommand{\feh} {\mbox{\rm [Fe/H]}}

\newcommand{\xh} {\mbox{\rm [X/H]}}

\newcommand{\oh} {\mbox{\rm [O/H]}}

\newcommand{\nah} {\mbox{\rm [Na/H]}}

\newcommand{\mgh} {\mbox{\rm [Mg/H]}}

\newcommand{\fracmgh} {\mbox{${\rm [\frac{Mg}{H}]}$}}
\newcommand{\fracfena} {\mbox{${\rm [\frac{Fe}{Na}]}$}}
\newcommand{\fracfemg} {\mbox{${\rm [\frac{Fe}{Mg}]}$}}
\newcommand{\fraccrmg} {\mbox{${\rm [\frac{Cr}{Mg}]}$}}

\newcommand{\sih} {\mbox{\rm [Si/H]}}

\newcommand{\cah} {\mbox{\rm [Ca/H]}}

\newcommand{\mnh} {\mbox{\rm [Mn/H]}}

\newcommand{\xfe} {\mbox{\rm [X/Fe]}}
\newcommand{\cfe} {\mbox{\rm [C/Fe]}}
\newcommand{\ofe} {\mbox{\rm [O/Fe]}}
\newcommand{\nafe} {\mbox{\rm [Na/Fe]}}
\newcommand{\mgfe} {\mbox{\rm [Mg/Fe]}}
\newcommand{\mgmn} {\mbox{\rm [Mg/Mn]}}
\newcommand{\alfe} {\mbox{\rm [Al/Fe]}}
\newcommand{\sife} {\mbox{\rm [Si/Fe]}}

\newcommand{\cafe} {\mbox{\rm [Ca/Fe]}}
\newcommand{\scfe} {\mbox{\rm [Sc/Fe]}}
\newcommand{\sch} {\mbox{\rm [Sc/H]}}
\newcommand{\tife} {\mbox{\rm [Ti/Fe]}}
\newcommand{\vfe} {\mbox{\rm [V/Fe]}}
\newcommand{\vh} {\mbox{\rm [V/H]}}
\newcommand{\crfe} {\mbox{\rm [Cr/Fe]}}
\newcommand{\mnfe} {\mbox{\rm [Mn/Fe]}}
\newcommand{\cofe} {\mbox{\rm [Co/Fe]}}
\newcommand{\coh} {\mbox{\rm [Co/H]}}
\newcommand{\coni} {\mbox{\rm [Co/Ni]}}
\newcommand{\nife} {\mbox{\rm [Ni/Fe]}}
\newcommand{\cufe} {\mbox{\rm [Cu/Fe]}}
\newcommand{\znfe} {\mbox{\rm [Zn/Fe]}}
\newcommand{\znni} {\mbox{\rm [Zn/Ni]}}

\newcommand{\xmg} {\mbox{\rm [X/Mg]}}

\newcommand{\almn} {\mbox{\rm [Al/Mn]}}

\newcommand{\crmg} {\mbox{\rm [Cr/Mg]}}

\newcommand{\femg} {\mbox{\rm [Fe/Mg]}}
\newcommand{\fena} {\mbox{\rm [Fe/Na]}}

\newcommand{\cumg} {\mbox{\rm [Cu/Mg]}}

\newcommand{\alphafe} {\mbox{\rm [$\alpha$/Fe]}}
\newcommand{\teff}  {\mbox{$T_{\rm eff}$}}

\newcommand{\logg}  {\mbox{{\rm log}\,$g$}}

\newcommand{\Msun}  {\mbox{$M_{\sun}$}}

\newcommand{\turb}  {\mbox{$\xi_{\rm turb}$}}

\newcommand{\CI} {\ion{C}{i}}
\newcommand{\OI} {\ion{O}{i}}

\newcommand{\MgI} {\ion{Mg}{i}}

\newcommand{\ScII} {\ion{Sc}{ii}}

\newcommand{\TiII} {\ion{Ti}{ii}}
\newcommand{\VI} {\ion{V}{i}}

\newcommand{\CrI} {\ion{Cr}{i}}
\newcommand{\CrII} {\ion{Cr}{ii}}

\newcommand{\FeI} {\ion{Fe}{i}}
\newcommand{\FeII} {\ion{Fe}{ii}}
\newcommand{\CoI} {\ion{Co}{i}}

\newcommand{\CuI} {\ion{Cu}{i}}

\newcommand{\VLSR}   {\mbox{$V_{\rm LSR}$}}
\newcommand{\ULSR}   {\mbox{$U_{\rm LSR}$}}
\newcommand{\WLSR}   {\mbox{$W_{\rm LSR}$}}

\def\ltsima{$\; \buildrel < \over \sim \;$}
\def\simlt{\lower.5ex\hbox{\ltsima}}
\def\gtsima{$\; \buildrel > \over \sim \;$}
\def\simgt{\lower.5ex\hbox{\gtsima}}
\begin{document}

\title{Abundances of iron-peak elements in accreted and in situ born Galactic halo stars
\thanks{Based on data products from observations made with ESO Telescopes
at the La Silla Paranal Observatory and the Nordic Optical Telescope
under programmes given in Tables 1 and 2 of Paper\,I.}
\fnmsep\thanks{Table A.1 is only available in electronic form at the CDS
via anonymous ftp to {\tt cdsarc.u-strasbg.fr (130.79.128.5a}) or via
{\tt http://cdsarc.u-strasbg.fr/cgi-bin/qcat?/A+A/XXX/xxx}.}}

\titlerunning{Abundances of iron-peak elements}

\author{P.E.~Nissen \inst{1} \and A.M.~Amarsi \inst{2} \and {\'A.}~Sk{\'u}lad{\'o}ttir \inst{3,4} \and W.J.~Schuster \inst{5} }

\institute{Department of Physics and Astronomy, Aarhus University, Ny Munkegade 120, DK--8000
Aarhus C, Denmark  \email{pen@phys.au.dk}
\and Theoretical Astrophysics, Department of Physics and Astronomy, Uppsala University, Box 516, SE--751 20 Uppsala, Sweden
\and Dipartimento di Fisica e Astronomia, Universit{\'a } degli Studi di Firenze, 
Via G. Sansone 1, I-50019 Sesto Fiorentino, Italy
\and INAF/Osservatorio Astrofisico di Arcetri, Largo E. Fermi 5, I-50125 Firenze, Italy
\and Instituto de Astronomia, Universidad Nacional Aut\'{o}noma
de M\'{e}xico, AP 106, Ensenada 22800, BC, M\'{e}xico}

\date{Received 27 October 2023  / Accepted 4 December 2023}

\abstract
{Studies of the element abundances and kinematics of stars belonging to the Galactic
halo have revealed the existence of two distinct populations: accreted stars with 
a low \alphafe\ ratio and in situ born stars with a higher ratio.} 
{Previous work on the abundances of C, O, Na, Mg, Si, Ca, Ti, Cr, Mn, Fe, Ni, Cu, and Zn
in high-$\alpha$ and low-$\alpha$ halo stars
is extended to include the abundances of Sc, V, and Co, enabling us to study the 
nucleosynthesis of all iron-peak elements along with the lighter elements.}
{The Sc, V, and Co abundances were determined from a 1D MARCS model-atmosphere analysis of equivalent widths 
of atomic lines in high signal-to-noise, high resolution spectra
assuming local thermodynamic equilibrium (LTE). 
In addition, new 3D and/or non-LTE calculations were used to correct the 1D LTE abundances
for several elements including consistent 3D non-LTE calculations for Mg.}
{The two populations of accreted and in situ born stars are well separated in 
diagrams showing \scfe , \vfe , and \cofe\ as a function of \feh .
The \xmg\ versus \mgh\ trends for high-$\alpha$ and low-$\alpha$ stars were used to determine the 
yields of core-collapse and Type Ia supernovae.
The largest Type Ia contribution occurs for Cr, Mn, and Fe, whereas Cu is a pure core-collapse
element. Sc, Ti, V, Co, Ni, and Zn represent intermediate cases.
A comparison with yields calculated for supernova models shows poor agreement for the
core-collapse yields. The Ia yields suggest that sub-Chandrasekhar-mass 
Type Ia supernovae provide a dominant contribution to the chemical evolution of the 
host galaxies of the low-$\alpha$ stars. A substructure in the abundances and kinematics
of the low-$\alpha$ stars suggests that they arise from at least two different 
satellite accretion events, {\it Gaia}-Sausage-Enceladus and Thamnos.} 
{}


\maketitle

\section{Introduction}
\label{introduction} 
In a study of the chemical composition of high-velocity stars, 
\citet[][hereafter Papers\,I and II]{nissen10, nissen11} found that F and G dwarf stars
with halo kinematics split into two distinct populations: $i$) one with
a high ratio between the abundance of  $\alpha$-capture elements and iron
($\alphafe \! \simeq \! 0.3$)\footnote{For two elements, X and Y,
with number densities $N_{\rm X}$ and $N_{\rm Y}$,
[X/Y] $\equiv {\rm log}(N_{\rm X}/N_{\rm Y})_{\rm star}\,\, - 
\,\,{\rm log}(N_{\rm X}/N_{\rm Y})_{\sun}$ and 
\alphafe\ is an unweighted mean of \mgfe , \sife , \cafe , and \tife .}
similar to stars with thick-disk kinematics; 
and $ii$) a population with $\alphafe$ decreasing from $\sim \! 0.3$\,dex
at a metallicity of $\feh = -1.6$ to $\alphafe \simeq 0.1$ at $\feh = -0.8$.  
A clear separation between the two populations was also
found for \nafe , \nife , \cufe , and \znfe , whereas \crfe\ and \mnfe\ overlap in their metallicity trends.
In a third paper \citep{schuster12}, the space velocities
of the stars were used to show that the  low-$\alpha$ stars move on radial orbits with maximum 
Galactocentric distances $r_{\rm max}$ up to 30-40\,kpc,
whereas the high-alpha stars move on more circular orbits with  $r_{\rm max}$ up to about 16\,kpc.
Furthermore, it was found that the low-$\alpha$ stars tend to be 2-3\,Gyr younger than 
the high-$\alpha$ halo stars. All of this was interpreted in a scenario in which the high-$\alpha$ stars
were formed in situ during the fast collapse of the proto-Galactic gas cloud with only 
core-collapse supernovae (CC SNe)
contributing to the chemical enrichment, whereas the low-$\alpha$ stars were accreted from dwarf
galaxies with a slower star formation rate allowing Type Ia supernovae (Ia SNe) to contribute
to the chemical evolution.

The existence of two discrete halo populations is supported by carbon and oxygen
abundances derived by \citet{amarsi19b} from a 3D model-atmosphere analysis of
equivalent widths of \CI , \OI , and \FeII\ lines taking departures
from local thermodynamic equilibrium (LTE) into account.
The dichotomy in halo star abundances is also confirmed
in the analysis of abundance data from the Apache Point Observatory Galactic Evolution
Experiment (APOGEE) for large samples of K giants by 
\citet{hawkins15} and \citet{hayes18}, who added \alfe\ and \almn\ as very useful indicators
of the separation between accreted and in situ born stars. We note that aluminium was not 
included in Papers I and II, because there are no suitable Al lines available in the
spectra used for the investigation.

With the advent of {\it Gaia} data, it was realised that halo stars in the solar neighbourhood 
show  an orbital anisotropy in the radial direction and an elongated structure
in the Toomre velocity diagram \citep{belokurov18, helmi18}. 
This has been interpreted as a signature of a merger of the Milky Way
with a massive dwarf galaxy named {\it Gaia}-Sausage-Enceladus (GSE) in which the majority
of the low-$\alpha$ stars 
were born \citep{helmi20}, but other merging dwarf galaxies have probably also contributed 
to the low-$\alpha$ population causing a cosmic scatter in 
\alphafe , \nafe , and \nife\ at a given \feh\ 
\citep[e.g.][]{myeong19, matsuno22a, matsuno22b, horta23, donlon23}. 
The high-$\alpha$ stars have been explained as being born in a precursor to
the Galactic thick disk and dynamically heated to halo kinematics due to the GSE merger,
the so-called big splash \citep{belokurov20}, although the more metal-poor part of the
high-$\alpha$ population may consist of stars born in situ from infalling gas before the formation of
a disk \citep{belokurov22, myeong22, feltzing23}.

In the last decade, there have also been several interesting studies of the relative ages of high-$\alpha$ and 
low-$\alpha$ stars. \citet{ge16} found the same age difference of about 2\,Gyr 
as \citet{schuster12} when using stellar models with C and O abundances from \citet{nissen14}.
A much larger sample of about 15\,000 stars in the turn-off region was analysed
by \citet{hawkins14}, who used Sloan Digital Sky Survey low-resolution spectra
to derive a spectral index of \alphafe .
From a comparison with isochrones, they found an age difference of about 1\,Gyr between high-$\alpha$ and
low-$\alpha$ stars in the metallicity range $-1.2 < \feh < -0.9$ but an equal age of the two
populations for more metal-poor stars. A similar metallicity trend of the age difference between in situ
and accreted globular clusters was found by \citet{massari19}. Furthermore, \citet{xiang22}
used precise age determinations of 150\,000 subgiant stars, based on estimates of
effective temperature \teff , absolute magnitude $M_K$, \feh , and \alphafe\
from the Large Sky Area Multi-Object fiber Spectroscopic Telescope (LAMOST) survey \citep{zhao12}, to show that
high-$\alpha$ halo stars are 1-2\,Gyr older than low-$\alpha$ stars
at a metallicity of $\feh \sim -1.0$. These results support the idea that the star formation
rate in the host galaxies of the low-$\alpha$ stars has been so slow that Ia SNe have significantly
contributed to the chemical evolution at low \feh\ \citep[e.g.][]{fernandez-alvar18, vincenzo19}, 
whereas the high-$\alpha$ stars are more likely to have been born
from gas that is enriched by only CC SNe. A particularly interesting modelling of the evolution
of abundance ratios in the GSE dwarf galaxy has been published by \citet{sanders21}; from the metallicity
trends of \mnfe\ and \nife , they find that sub-Chandrasekhar-mass (sub-\Ch )
Ia SNe provide a significant contribution to the chemical evolution.

With these recent results in mind, we decided to have a fresh look at the abundances
for the high-$\alpha$ and low-$\alpha$ populations in Papers I and II and
to add abundances of Sc, V, and Co. Thus, we can  provide the most
precise and accurate iron-peak abundances available for in situ and ex situ halo stars,
allowing important tests of star formation and nucleosynthesis.

\section{Sc, V, and Co abundances}
\label{abundances}
In Papers I and II, abundances of Na, Mg, Si, Ca, Ti, Cr, Mn, Fe, Ni, Cu, and Zn 
were determined for 94 main-sequence stars selected from  Str\"{o}mgren photometry 
\citep{schuster06} to have metallicities in the range $-1.6 < \feh < -0.4$ and 
\teff\ from 5300\,K to 6400\,K.
\citet{fishlock17} have determined Sc abundances for 27 of these stars,
but using only the \ScII\ line at 5526.8\,\AA .
We have used eight \ScII\ lines (see Table\,\ref{table:linedata}) and derived Sc abundances
for 85 stars, that is, those included in the papers on C and O abundances by
\citet{nissen14} and \citet{amarsi19b}. In addition, V and Co abundances were determined from a set of four
lines for each element as listed in Table\,\ref{table:linedata}. These lines are too weak 
to be measured for the warmer and more metal-poor stars in our sample,   
but we were able to derive reliable V and Co abundances for
60 and 59 stars, respectively.

\begin{figure}
\centering
\resizebox{\hsize}{!}{\includegraphics{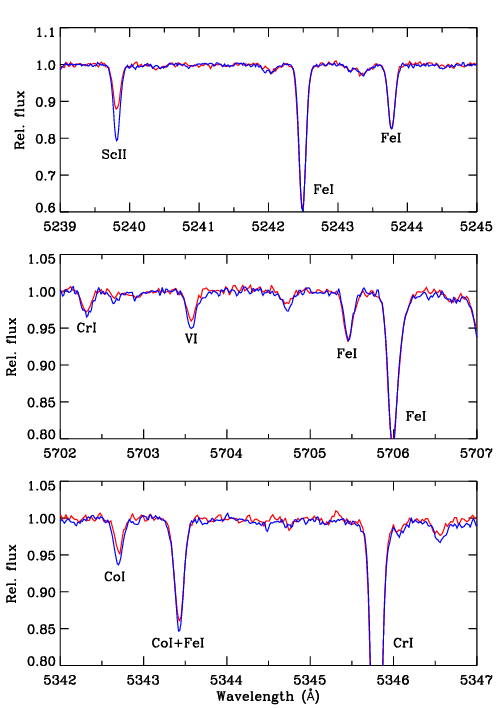}}
\caption{Spectra near representative Sc, V, and Co lines
for a high-$\alpha$ and a low-$\alpha$ star with similar \teff , \logg , and \feh\ values.
The spectrum of
\object{G\,159-50} (\teff \,=\,5713\,K, \logg \,=\,4.44,
\feh\,=\,$-0.94$, \alphafe \,=\,0.32) is shown with a blue line, and that of
\object{CD\,$-45\,3283$} (\teff \,=\,5685\,K, \logg \,=\,4.61,
\feh \,=\,$-0.93$, \alphafe \,=\,0.12) with a red line.
}
\label{fig:spectra}
\end{figure}

Figure \ref{fig:spectra} shows a comparison of spectra near some typical \ScII , \VI ,
and \CoI\ lines for a high-$\alpha$ and a low-$\alpha$ star
with similar \teff , logarithmic surface gravity \logg , and \feh , but a difference of 0.20\,dex in \alphafe .
As seen, the two stars have the same strengths of \CrI\ and \FeI\ lines, whereas the \ScII\ line
is about 60\% stronger in the high-$\alpha$ star (a small part of this is due to the difference
in \logg ). The \VI\ and \CoI\ lines are also stronger in the 
high-$\alpha$ star, but the differences are not so large, 
and due to the weakness of these lines, it is more difficult to derive
precise values of the differences in V and Co abundances.

The IRAF {\tt splot} task was used to
measure equivalent widths (EWs) of the Sc, V, and Co lines by Gaussian fitting
to the line profiles in the spectra from Paper I.
During the measurements, it was discovered that two low-$\alpha$ stars, HD\,163810 and HD\,250792\,A,
have somewhat asymmetric line profiles, clearly seen for the low-excitation \VI\ lines.
According to the Washington Visual Double Star Catalogue \citep{mason20}
both stars have 1 - 2 magnitudes fainter components within 0.5 arcsec from the primary component,
and because the derived \vfe\ ratios are unusually high (about 0.20\,dex higher than 
expected from the \vfe \,-\,\feh\ relation for the other low-$\alpha$ stars), we have
excluded the two stars from the discussion of abundances.

\begin{table}
\caption[ ]{Spectral line data and equivalent widths.}
\label{table:linedata}
\setlength{\tabcolsep}{0.18cm}
\begin{tabular}{rcccrrc}
\noalign{\smallskip}
\hline\hline
\noalign{\smallskip}
  ID & Wavelength & $\chi_{\rm exc}$ & log($gf$)\,\tablefootmark{a} & EW\,\tablefootmark{b} & EW\,\tablefootmark{c} & Ref.  \\
          &  (\AA )    &  (eV)  &   &  (m\AA )  & (m\AA )  & HFS  \\
\noalign{\smallskip}
\hline
\noalign{\smallskip}
ScII &      5239.82  & 1.46 &  $-0.75$  & 29.7 &  31.5 & 1 \\  
ScII &      5526.82  & 1.77 &  $-0.05$  & 52.9 &  55.9 & 1 \\ 
ScII &      5657.88  & 1.51 &  $-0.49$  & 41.7 &  44.1 & 1 \\
ScII &      5667.15  & 1.50 &  $-1.19$  & 13.7 &  13.9 & 1 \\
ScII &      5669.04  & 1.50 &  $-1.07$  & 16.7 &  17.8 & 1 \\
ScII &      5684.20  & 1.51 &  $-1.00$  & 18.4 &  20.4 & 1 \\
ScII &      6245.62  & 1.51 &  $-1.11$  & 16.0 &  17.6 & 2 \\
ScII &      6604.60  & 1.36 &  $-1.30$  & 15.6 &  16.3 & 1 \\
 VI  &      4875.49  & 0.04 &  $-0.81$  & 12.3 &   9.7 & 3 \\
 VI  &      5703.59  & 1.05 &  $-0.20$  &  5.8 &   5.3 & 3 \\
 VI  &      5727.06  & 1.08 &    +0.02  &  8.4 &   8.7 & 3 \\
 VI  &      6090.22  & 1.08 &  $-0.05$  &  7.7 &   7.4 & 3 \\
CoI  &      5342.71  & 4.02 &    +0.54  &  7.7 &   7.3 & 2 \\
CoI  &      5352.05  & 3.58 &  $-0.10$  &  5.2 &   4.4 & 2 \\
CoI  &      5369.60  & 1.74 &  $-1.47$  & 11.6 &  10.3 & 2 \\ 
CoI  &      5483.36  & 1.71 &  $-1.49$  & 12.1 &  11.1 & 4 \\
\noalign{\smallskip}
\hline
\end{tabular}
\tablefoot{
\tablefoottext{a} {Astrophysically calibrated as explained in the text;}
\tablefoottext{b} {Equivalent widths for HD\,22879;}
\tablefoottext{c} {Equivalent widths for HD\,76932.}
}

\tablebib{
(1)~\citet{lawler19}; (2)~\citet{prochaska00}; (3)~\citet{Lawler14}; (4)~\citet{Lawler15}.
}
\end{table}

The Uppsala BSYN code was used to calculate EWs as a function of element abundance
assuming LTE and taking hyper-fine splitting (HFS)
into account according to the references given in Table\,1. Interpolation to the observed EW
then yields the element abundance. First, two bright standard stars,  HD\,22879 and
HD\,76932, with well known photometric values of \teff\ and \logg\ were analysed differentially with respect 
to the Sun avoiding saturated lines or lines significantly blended in the 
solar flux spectrum. Adopting the derived mean \xh\ abundances for the standard stars
and solar abundances $A$(Sc) = 3.14, $A$(V) = 3.90, 
and $A$(Co) = 4.94\,\footnote{$A$(X)  = log\,($N_{\rm X} / N_{\rm H}$) + 12.0}
from \citet{asplund21}, an `inverted' abundance analysis of  HD\,22879 and HD\,76932
then yields the log\,$gf$ value of all lines as given in Table\,1. With these 
log\,$gf$ values, the abundances of the programme stars were derived.

\begin{figure*}
\centering
\includegraphics[width=16.0cm]{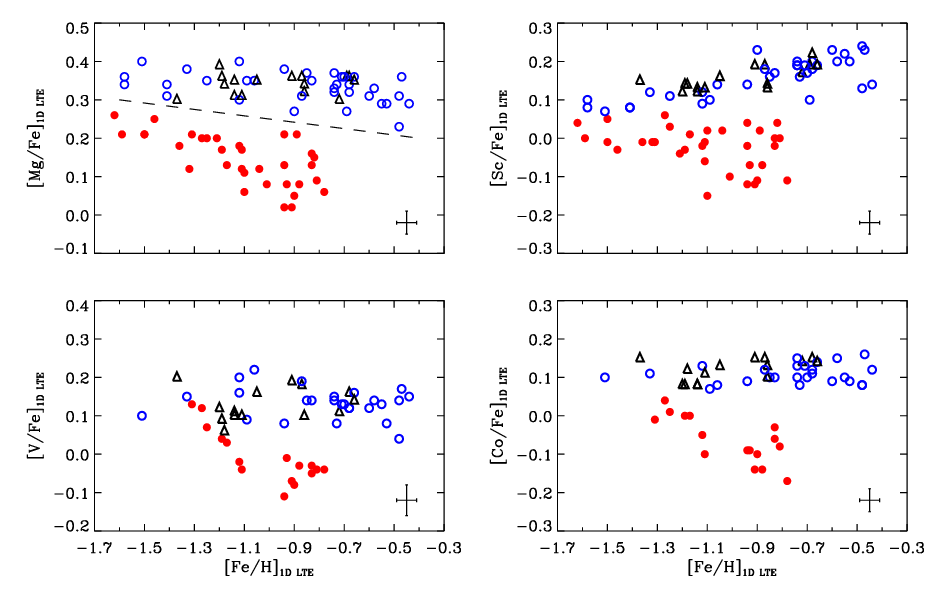}
\caption{\mgfe , \scfe , \vfe , and \cofe\ versus \feh .
Stars with halo kinematics are classified from the \mgfe\,-\,\feh\ diagram as belonging
to either the high-$\alpha$ (blue circles) or the low-$\alpha$ (filled red circles) population.
Stars with thick-disk kinematics are labelled with black triangles. The dashed line in the
\mgfe\,-\,\feh\ diagram indicates the separation between high-$\alpha$ and low-$\alpha$ stars adopted in
Paper\,I.}
\label{fig:all-feh}
\end{figure*}

Model atmospheres were interpolated from the plane parallel (1D) MARCS grid \citep{gustafsson08} to the 
spectroscopic \teff , \logg , \feh , and \alphafe\ values determined in \citet{nissen14}.
The \teff\ values are about 100\,K
higher than those in Papers I and II, because revised photometric temperatures of
the standard stars were adopted from \citet{casagrande10}. With this change of the
\teff\ scale and the corresponding changes of \logg , we have also updated 
the abundances derived in Papers I and II, but the changes
of \xfe\ are within $\pm 0.02$\,dex.

From the line-to-line abundance scatter and the changes in derived abundances, when changing
the atmospheric parameters and the microturbulence \turb\ with one-sigma uncertainties
($\sigma \teff \! = \! \pm 35$\,K, $\sigma \logg \! = \! \pm 0.05$\,dex, and
$\sigma \turb \! = \pm \! 0.05$\,\kmprs ), we estimate that the
one-sigma error is $\pm 0.03$\,dex for \scfe\ and \cofe\ and 
$\pm 0.04$\,dex for \vfe . The errors of abundance ratios with respect to Mg are about 0.01\,dex higher.
We emphasise that these numbers are internal statistical errors; in addition,
there may be systematic errors in the derived abundance ratios due to the
adoption of 1D model atmospheres and the assumption of LTE as discussed in Sect. 4.1.

The derived values of \sch , \vh , and \coh\ are given in the online Table A.1 together with
the atmospheric parameters of the stars.

\section{Separation of accreted and in situ born stars}
\label{abundance ratios}
Figure \ref{fig:all-feh} shows \mgfe\ together with \scfe , \vfe , and \cofe\
as a function of \feh . Using the same classification as in Paper\,I, 
based on the \mgfe\,-\,\feh\ diagram and the kinematics of the stars, we have divided
stars with halo kinematics, that is, those having a total space velocity with respect 
to the Local Standard of Rest (LSR), $\VLSR > 180$\,\kmprs , into 
high-$\alpha$ and low-$\alpha$ populations. They are shown with blue and red circles, respectively. 
Stars with typical thick-disk kinematics are shown with black triangles.

\begin{figure*}
\centering
\includegraphics[width=16.0cm]{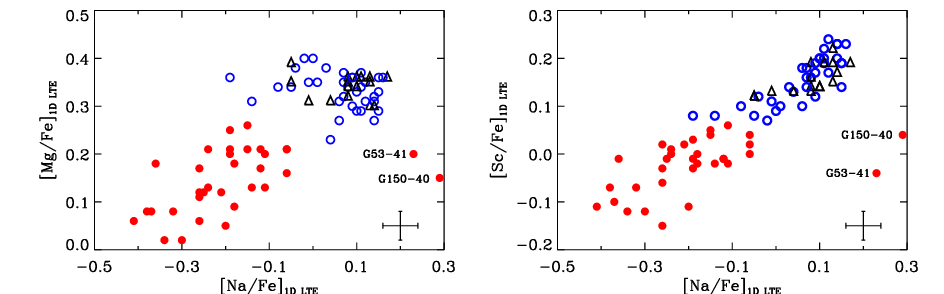}
\caption{\mgfe\ and \scfe\ as a function of \nafe\ with the same symbols as in 
Fig. \ref{fig:all-feh}.}
\label{fig:mgsc-na}
\end{figure*}

As seen from Fig. \ref{fig:all-feh}, the high-$\alpha$ halo stars and the thick-disk stars
have well defined metallicity trends with no systematic offset between the two populations.
The standard deviation relative to fitted second order polynomials is 0.033\,dex for \mgfe , 0.030\,dex for \scfe ,
0.039\,dex for \vfe , and 0.026\,dex for \cofe . This is close to the estimated one-sigma errors of
the abundance ratios. Hence, there is no evidence of a cosmic scatter at a given \feh\ for the
high-$\alpha$ population. This supports that the high-$\alpha$ halo and the thick-disk
stars were formed in regions with a high star formation rate over a short time such that Ia SNe did not
contribute to the chemical evolution. 

While \mgfe\ is nearly constant for the  high-$\alpha$ 
population, \scfe\ shows a rising trend with increasing \feh . This suggests that Sc has 
metallicity dependent CC SNe yields. The two other odd-$Z$
elements in  Fig. \ref{fig:all-feh}, vanadium ($Z = 23$) and cobalt ($Z = 27$) have, however, nearly
constant \xfe\ ratios as a function of \feh . This is in agreement with the yield calculations for
CC SNe by \citet{kobayashi06}. According to their Table 3, the initial mass function (IMF) weighted
yield of Sc increases with a factor of two from $\feh \simeq -1.3$ to $\feh \simeq -0.7$, whereas the yields of 
V and Co do not change significantly over the same metallicity range. 

The increase in \scfe\ for the high-$\alpha$ stars amounts to $\sim \! 0.15$\,dex, 
when \feh\ increases from $-1.6$ to $-0.5$. This gradient of \scfe\ refers to 1D LTE abundances.
In a recent work,
\citet{mashonkina22} have shown that non-LTE corrections of Sc abundances derived from \ScII\ lines 
are significant. For a sample of 56 main-sequence stars in the same \teff\ range as our stars, they
find that the average non-LTE correction increases by $\sim \! 0.08$\,dex, when \feh\ decreases from 
$-0.5$ to $-1.6$\,dex \citep[see Table 2 in][]{mashonkina22}. If such non-LTE corrections are applied 
to our derived Sc abundances, the gradient of \scfe\ is reduced by $\sim \! 50$\%.
Clearly, it would be interesting to make a non-LTE study of the \ScII\ lines in Table\,\ref{table:linedata}
for our set of stars, preferably with 3D model atmospheres. Studies of non-LTE corrections for V and Co
are also needed.

Figure \ref{fig:all-feh} shows that the low-$\alpha$ population, defined as stars falling below the dashed line 
in the \mgfe\,-\,\feh\ diagram, is also well separated from the high-$\alpha$ population in the
\scfe\,-\,\feh\ diagram, although the separation is a bit smaller
than the separation in \mgfe .  
The two populations are also well separated in \vfe\ and \cofe\ in the 
metallicity range $-1.1 < \feh < -0.7$, but for these two elements, the two sequences
tend to merge at $\feh \simeq -1.3$, whereas this happens at $\feh \simeq -1.6$ for
\scfe\ and at an even lower metallicity in the case of \mgfe .

As discussed in Paper II, there is a significant cosmic scatter in \xfe\ among
low-$\alpha$ stars with $\feh \ge -1.1$, in particular
for \nafe , \nife , and \cufe , for which the rms scatter in \xfe\ is a factor $\sim\!3$ higher than
the corresponding scatter for the high-$\alpha$ population (see Table 5 in Paper II).
This can be explained if several merging dwarf galaxies with different
star formation rates  have contributed to the low-$\alpha$ accreted population.
In the case of Sc, the rms scatter in \scfe\ for the low-$\alpha$ stars 
with $\feh \ge -1.1$ is 0.064\,dex, that is, a factor $\sim\!2$ higher than the 
scatter in \scfe\ for the high-$\alpha$ stars at a given \feh . For V and Co there are
too few low-$\alpha$ stars to get a precise estimate of the dispersion in \vfe\ and \cofe . 

Based on abundance data from the APOGEE survey, diagrams with \alphafe\ or \mgmn\ as a function of \alfe\
have proven to be very useful in separating accreted stars from in situ formed stars
\citep[e.g.][]{hawkins15, feuillet21, limberg22, fernandes23}. 
As mentioned above, we have not measured \alfe , but one may instead use \nafe\ 
as shown in Paper II and by \citet{buder22} when selecting accreted stars
in the Galactic Archaeology with HERMES (GALAH) survey.
As seen from Fig. \ref{fig:mgsc-na}, the \scfe\,-\,\nafe\ diagram can also be used for this purpose. Except for
a small overlap among the more metal-poor ($\feh < -1.4$) stars, 
high-$\alpha$ and low-$\alpha$ stars are well separated.
Furthermore, there is a remarkably tight correlation between \scfe\ and \nafe\ with the exception
of two Na-rich stars, \object{G\,53-41} and \object{G\,150-40}. These two stars have also
exceptional low C and O abundances  \citep{ramirez12, nissen14} suggesting that they 
are second generation globular cluster escapees enriched in Na made through the
CNO and Ne-Na cycles of hydrogen burning. Possible first-generation polluters are
intermediate-mass AGB stars \citep{ventura01} or fast-rotating massive stars
\citep{decressin07}.

\section{3D and non-LTE effects}
\label{3DNLTE}
In Sect. \ref{nucleosynthesis}, the chemical evolution and nucleosynthesis of 
elements for the high-$\alpha$ and
low-$\alpha$  populations are discussed based on trends in \xmg \,-\,\mgh\ diagrams.
In this connection, consideration of possible 
3D non-LTE effects is important, in particular for magnesium,
because it is the reference element. We have, therefore, applied
3D non-LTE abundance corrections for the \MgI\ 5711.1\,\AA\ line, which is the only line used
to derive Mg abundances in this paper\,\footnote{In Paper I, the \MgI\ 4730.2\,\AA\ line 
was included for stars observed with the NOT/FIES spectrograph.}. 
The full details will be presented in a future study
(Amarsi et al. in prep.). In brief, the calculations were carried out using \texttt{Balder} 
\citep{amarsi18, amarsi22}, which is a modified version of \texttt{Multi3D} 
\citep{leenaarts09}. The model atom was that presented in  \citet{asplund21}. First, 1D non-LTE
corrections were computed on a fine grid of \texttt{MARCS} models, a subset of those used in
\citet{amarsi20} that covers the stellar parameters studied here. Secondly, 3D non-LTE
versus 1D non-LTE corrections were applied to these based on calculations on a coarse grid of
1D and 3D models from the \texttt{Stagger}-grid \citep{magic13}, spanning 5000\,K to
6500\,K in $T_{\mathrm{eff}}$, 4.0 to 5.0 in $\log g$, $-3.0$ to $0.0$ in [Fe/H], and $-0.8$ to
$0.8$ in [Mg/Fe]; the 1D calculations were performed for three values of microturbulence,
namely $0$, $1$, and $2\,\mathrm{km\,s^{-1}}$.

The corrections of \mgh\ range from $-0.03$\,dex
for the coolest stars to $+0.07$\,dex for the warmest and most metal-poor stars. 
We stress that these are differential corrections: the 3D non-LTE correction for the Sun
(+0.057\,dex) has been subtracted out. The corrections have
some effect on the slopes of the \xmg \,-\, \mgh\ relations, which are  
increased by  0.05 to 0.10 relative to the values obtained
without the 3D non-LTE corrections. At a given \mgh , the difference
in \xmg\ between the low-$\alpha$ and high-$\alpha$ stars 
is, however, not changed significantly.

For the C and O abundances, we have used the 3D non-LTE values presented in  \citet{amarsi19b}.
For Na, Si, and Ca, new 1D non-LTE calculations were 
performed with \texttt{Balder} on a fine grid of \texttt{MARCS} models covering the stellar
parameters and spectral lines studied here\,\footnote{Table 3 in Paper II contains a list of all
spectral lines used to determine abundances.}. The sodium atom was that presented in \citet{lind11} and the
calcium atom that presented in \citet{asplund21}; they were recently also applied in
\citet{barklem21} and \citet{skuladottir21}, respectively. The silicon atom was based on
that used in \citet{amarsi17}. Here, the atom was updated to use inelastic hydrogen
collisions based on the asymptotic model of \citet{barklem16} combined with the free electron
model of \citet{kaulakys91} in the scattering length approximation as described in
\citet{amarsi18}. The reduced model atom is also more complex than before due to the inclusion of fine
structure levels up to $7\,\mathrm{eV}$ above the ground state.

While the differential 3D non-LTE corrections of \oh\ derived from the \OI\ 7774\,\AA\ triplet
range from $-0.05$ to $+0.10$\,dex for our sample of stars, the 1D non-LTE corrections
of \nah , \sih , and \cah\ are less significant. The differential corrections 
range from $-0.03$ to +0.04\,dex for \nah , from $-0.01$ to +0.01\,dex for \sih ,
and from 0.00 to +0.07\,dex for \cah .  

Furthermore, we have have applied 1D non-LTE corrections for Mn from
\citet{bergemann19} using {\tt Spectrum Tools}\,\footnote{{\tt http://nlte.mpia.de}.} to 
interpolate to the atmospheric parameters of our stars. 
The  metallicity trend of \mnfe\ 
is very sensitive to these corrections; \citet{eitner20} found that the
\mnfe \,-\,\feh\ trend  becomes nearly flat when applying the \citet{bergemann19} corrections.
In our case, the differential non-LTE corrections of \mnh\ range from
$\sim \! 0.0$\,dex for the coolest stars to about +0.20\,dex for the warmer and most metal-poor stars 
in the sample. 

For Cu, we have adopted the abundances derived in the
1D non-LTE re-analysis of \CuI\ lines from Paper II by \citet{yan16}. 
As in the case of Mn, this leads to a less steep
slope of the \cumg \,-\mgh\ relation than obtained with LTE abundances.

As a number of suitable \TiII , \CrII , and \FeII\ lines are available in our spectra,
we have used these lines to derive abundances. This has the advantage that non-LTE 
effects are estimated to be insignificant
\citep{bergemann10, bergemann11, lind12}. In the case of Fe, we used 3D abundances from
\citet{amarsi19b}, but the 3D\,--\,1D difference have an rms scatter of only 0.017\,dex.
The Ti and Cr abundances were derived from 1D models.

For the remaining elements (Sc, V, Co, Ni and Zn) we do not have 3D and/or non-LTE
corrections available for the spectral lines used to determine abundances. 
To the extent that the corrections depend on metallicity,
the slopes of the \xmg \,-\,\mgh\ relations may, therefore, be incorrect
(see discussion of Sc in Sect. \ref{abundance ratios}).
For all the elements discussed above, the mean difference in \xmg\
between the low-$\alpha$ and the high-$\alpha$ stars is, however,
practically independent on whether 3D and/or non-LTE corrections are applied or not. 
This is because the average \teff\ and \logg\ values of the two populations 
are about the same. 

The final abundances are given in 
the online Table A.1. This table also includes the 
3D non-LTE corrections for Mg and the 1D non-LTE corrections for Na, Si, Ca, and Mn abundances. 
The 3D non-LTE corrections for C and O abundances and the 3D corrections for Fe abundances derived from 
\FeII\ lines may be obtained from \citet{amarsi19b} and the 1D non-LTE corrections for Cu
abundances from \citet{yan16}.

\begin{figure*}
\centering
\includegraphics[width=18.0cm]{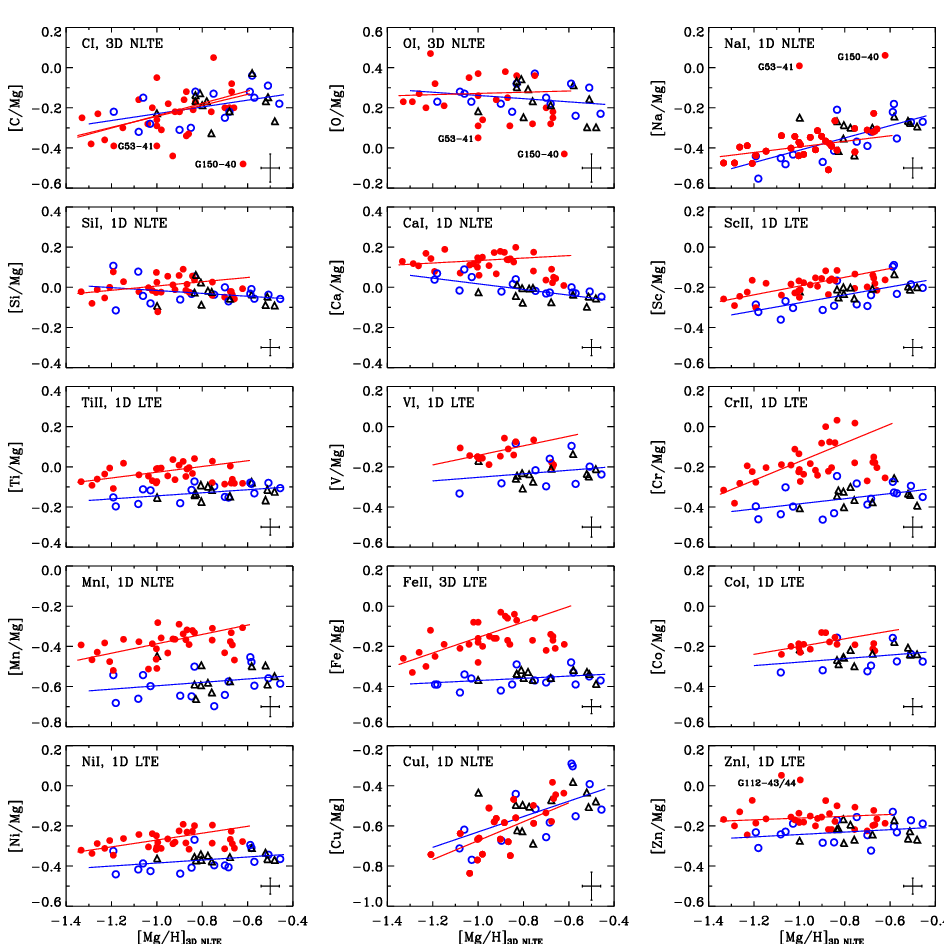}
\caption{\xmg\ as a function of \mgh\ with the same symbols as in
Fig. \ref{fig:all-feh}. The blue lines show least-squares fits to the high-$\alpha$ stars
and the red lines show fits to the low-$\alpha$ stars after excluding stars 
probably associated with the Thamnos substructure (see Fig. \ref{fig:fe2-all}).
On each panel, the atomic species of spectral lines used for deriving abundances
and the method of analysis are indicated.}
\label{fig:xmg-mgh}
\end{figure*}

\section{Nucleosynthesis of the elements}
\label{nucleosynthesis}
Predictions from Galactic chemical evolution (GCE) models are normally compared to observed
\xfe \,-\feh\ relations. The choice of Fe as a reference element is connected to the
fact that iron has far the largest set of lines in spectra of late-type stars
and is therefore usually the element with the most precise stellar abundances.
Iron has, however, significant yield contributions from both CC and Ia SNe,
which complicates the interpretation of the \xfe \,-\feh\ relations. As shown
by \citet{weinberg19, weinberg22} based on APOGEE abundances, it is more straightforward
to use a reference element that is produced almost
entirely in CC SNe, such as O or Mg. They found that the thick and the thin disk populations
in the Milky Way have different \xmg\ versus \mgh\ relations, but for each population the relation
is nearly constant throughout the disk. This is not the case
for the trends of \xfe\ versus \feh .

Following the approach of \citet{weinberg19}, we show in
Fig. \ref{fig:xmg-mgh} the \xmg \,-\mgh\ relations for high-$\alpha$ and low-$\alpha$ stars
in the $-1.4 < \mgh < -0.4$ range, where the two populations overlap. 
As seen, there is no significant difference in the 
\xmg \,-\mgh\ relations of the high-$\alpha$ and low-$\alpha$ populations for
C, O, Na, and Cu, whereas the largest difference between the two populations is seen for 
Cr, Mn, and Fe. This suggests that C, O, Na, and Cu have no contribution from Ia SNe, 
whereas Cr, Mn, and Fe get the largest Ia contribution.
The remaining elements represent intermediate cases.  In Sect. \ref{yields}, 
we present a more detailed investigation of this scenario, but first we
discuss the possible existence of substructure in the \xmg\,-\,\mgh\ relations 
for the low-$\alpha$ population.

\begin{figure*}
\centering
\includegraphics[width=16.0cm]{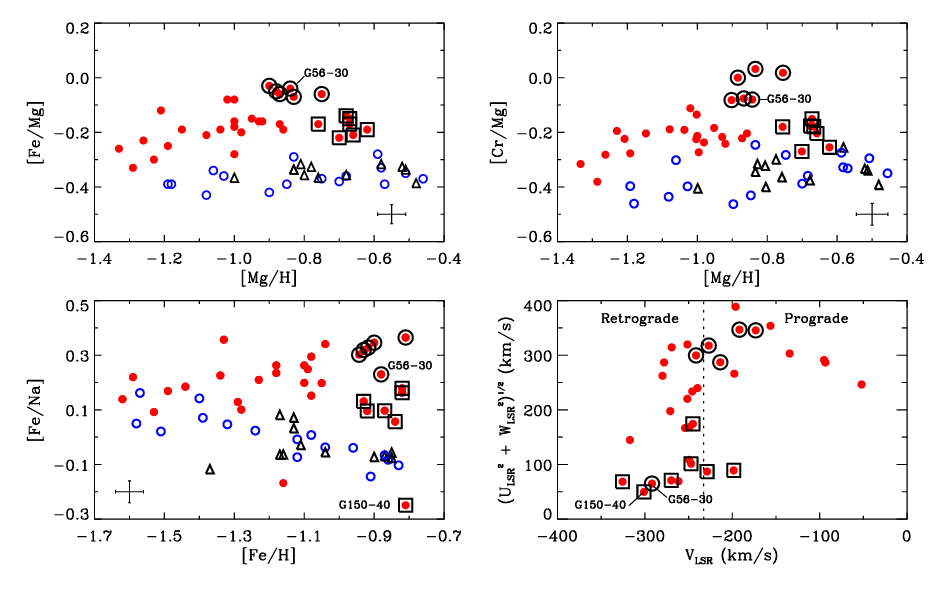}
\caption{Substructure in the distribution of \femg , \crmg , and \fena\
for low-$\alpha$ stars with $\mgh \simeq -0.8$. Stars with high \femg\ values are labelled with
black circles and those with lower \femg\ values with squares. 
In the lower right panel, the Toomre diagram for the low-$\alpha$ stars is
shown with the same labelling.} 
\label{fig:fe2-all}
\end{figure*}

\subsection{Substructure in the low-$\alpha$ population}
\label{substructure}
As mentioned in Sect. \ref{abundance ratios}, there is a significant cosmic scatter 
in \nafe , \scfe , \nife , and \cufe\ among the low-$\alpha$ stars.
The dispersion in some of the abundance ratios shown in Fig. \ref{fig:xmg-mgh}
is also higher than expected from the measurement error, most notable in the case
of \femg\ for which the standard deviation around the linear fit is 0.070\,dex for the
low-$\alpha$ population compared to 0.033\,dex for the high-$\alpha$ stars.
The main contribution to this higher dispersion 
comes from a splitting of stars with $\mgh \sim -0.8$ into two groups,
that is, six high-\femg\ stars
labelled with circles in Fig. \ref{fig:fe2-all} and seven low-\femg\ stars 
labelled with squares (see list in Table \ref{high.low.Fe.stars}).
The same distribution is present for \crmg\ as seen in the upper right panel of  Fig. \ref{fig:fe2-all}. 
Interestingly, this substructure is only clearly seen after applying the 
3D non-LTE corrections for Mg \citep{lind24}.

It could be that the substructure in  \femg\ and \crmg\ is due to 
errors in the Mg abundances, but a similar substructure is seen 
if we replace Mg with Na as the reference element 
and use \feh\ instead of \mgh\ on the abscissa (lower left panel of Fig. \ref{fig:fe2-all})
except that one high-\femg\ star, G56-30, falls between the two groups, and the Na-rich star,
G150-40, is shifted to a very low value of \fena .  

\begin{table}
\caption[ ]{Abundance ratios and kinematics of high-\femg\ and low-\femg\ stars in the 
low-$\alpha$ population.}
\label{high.low.Fe.stars}
\setlength{\tabcolsep}{0.09cm}
\begin{tabular}{crrrrrrr}
\noalign{\smallskip}
\hline\hline
\noalign{\smallskip}
  ID & \fracmgh & \fracfemg & \fraccrmg & \fracfena & \ULSR\ & \VLSR\ & \WLSR\  \\
     &   &   &   &  & km/s & km/s  & km/s \\
\noalign{\smallskip}
\hline
\noalign{\smallskip}
 &   \multicolumn{6}{c}{high-\femg\ stars} \\
\noalign{\smallskip}
       CD$-45\,3283$ &   $-$0.88 &   $-$0.05 &   $-$0.00  &     0.32 &  $-$302 &  $-$227 &   $-$98 \\
       CD$-57\,1633$ &   $-$0.87 &   $-$0.06 &   $-$0.07  &     0.32 &  $-$298 &  $-$241 &   $-$32 \\
            G\,56-30 &   $-$0.84 &   $-$0.04 &   $-$0.08  &     0.23 &      35 &  $-$292 &   $-$55 \\
            G\,66-22 &   $-$0.83 &   $-$0.07 &      0.03  &     0.35 &  $-$285 &  $-$214 &      31 \\
            G\,82-05 &   $-$0.75 &   $-$0.06 &      0.01  &     0.37 &  $-$343 &  $-$173 &      40 \\
           G\,121-12 &   $-$0.90 &   $-$0.03 &   $-$0.08  &     0.31 &  $-$330 &  $-$192 &     106 \\
\noalign{\smallskip}
 &   \multicolumn{6}{c}{low$-$\femg\ stars} \\
\noalign{\smallskip}
            G\,46-31 &   $-$0.67 &   $-$0.15 &   $-$0.18  &     0.16 &      32 &  $-$326 &      61 \\
            G\,56-36 &   $-$0.70 &   $-$0.22 &   $-$0.27  &     0.09 &  $-$101 &  $-$247 &       8 \\
            G\,98-53 &   $-$0.66 &   $-$0.21 &   $-$0.20  &     0.09 &  $-$151 &  $-$245 &   $-$87 \\
           G\,150-40 &   $-$0.62 &   $-$0.19 &   $-$0.26  &  $-$0.25 &   $-$49 &  $-$301 &      11 \\
           G\,170-56 &   $-$0.76 &   $-$0.17 &   $-$0.18  &     0.13 &   $-$61 &  $-$270 &   $-$36 \\
          HD\,103723 &   $-$0.68 &   $-$0.14 &   $-$0.17  &     0.18 &   $-$70 &  $-$198 &      55 \\
          HD\,105004 &   $-$0.67 &   $-$0.17 &   $-$0.15  &     0.06 &   $-$36 &  $-$229 &   $-$79 \\
\noalign{\smallskip}
\hline
\end{tabular}
\end{table}

In order to see if the substructure in the abundance diagrams of Fig. \ref{fig:fe2-all}
is related to the kinematics of the stars, we have plotted all low-$\alpha$ stars
in the Toomre velocity diagram shown in the lower right panel of Fig. \ref{fig:fe2-all}.
The space velocities with
respect to the LSR were calculated based on {\it Gaia} DR3 data as described in detail
by \citet{nissen21}; typical errors of the velocity components are 1-2\,\kmprs .

Five of the six high-\femg\ stars fall in the same high-energy region of the Toomre diagram as
stars accreted from the GSE dwarf galaxy \citep{belokurov18, helmi18}; they have large (negative) 
values of \ULSR\ (see Table \ref{high.low.Fe.stars}) and \VLSR\ close to the value 
dividing prograde from retrograde moving stars, $\VLSR = -233$\,\kmprs  \citep{mcmillan17}.   
The exception is G\,56-30, the star with an intermediate value of \fena .
Six of the seven low-\femg\ stars are, on the other hand, situated in the low-energy part
of the Toomre diagram and the majority are retrograde moving. We suggest that they
belong to the Thamnos substructure identified by \citet{koppelman19a}.
Interestingly, both \citet{koppelman19a} and \citet{horta23} find evidence from APOGEE abundances that 
Thamnos stars have \mgfe\ between the mean values of \mgfe\ for GSE and high-$\alpha$ stars, respectively,
in agreement with our results. Hence, both the abundance ratios and the kinematics point
to the existence of two groups of accreted stars among our more metal-rich low-$\alpha$ stars,
that is, debris stars from the GSE and Thamnos dwarf galaxies. 

The abundance ratios shown in Fig. \ref{fig:fe2-all} suggest that the contribution
from Ia SNe to the chemical evolution was more important at a given \mgh\
in the GSE galaxy than in the Thamnos dwarf galaxy. When discussing the nucleosynthesis of
elements in the following sections this will be taken into account. 
Therefore, Fig. \ref{fig:xmg-mgh} shows linear least-squares fits to the 
low-$\alpha$ stars after excluding the seven Thamnos stars.
Furthermore, the two C- and O-poor, Na-rich stars in the C, O, and Na panels
and the Zn-rich binary star G\,112-43/44 in the Zn panel \citep{nissen21}
have been excluded from the fits. Coefficients for the fits are given in Table \ref{table:fits}.

\begin{table*}
\centering
\caption[ ]{Coefficients of the linear fits, $\xmg = a + b\, (\mgh + 0.8)$, shown in Fig. \ref{fig:xmg-mgh},
and the derived  CC and Ia SNe yields.}
\label{table:fits}
\setlength{\tabcolsep}{0.20cm}
\begin{tabular}{cccccccc}
\noalign{\smallskip}
\hline\hline
\noalign{\smallskip}
   & \multicolumn{2}{c}{high-$\alpha$ stars} & \multicolumn{2}{c}{low-$\alpha$ stars} &   &   &   \\
  X & $a$ & $b$ &  $a$ & $b$ &  $F_{\rm X}$ & $\xfe _{\rm CC}$ &  $\xfe _{\rm Ia}$    \\
\noalign{\smallskip}
\hline
\noalign{\smallskip}
 C  & $-0.196 \pm 0.015$ & $+0.17 \pm 0.07$ & $-0.178 \pm 0.035$ & $+0.31 \pm 0.13$ &                 &                 & \\
 O  & $+0.245 \pm 0.015$ & $-0.08 \pm 0.07$ & $+0.277 \pm 0.042$ & $+0.03 \pm 0.14$ & $0.08 \pm 0.13$ & $+0.60 \pm 0.06$ & \\
 Na & $-0.350 \pm 0.013$ & $+0.31 \pm 0.06$ & $-0.367 \pm 0.018$ & $+0.14 \pm 0.07$ &                 &                 & \\
 Mg &                    &                  &                    &                  &    0.00         & $+0.36 \pm 0.06$ & \\
 Si & $-0.030 \pm 0.010$ & $-0.07 \pm 0.05$ & $+0.028 \pm 0.015$ & $+0.11 \pm 0.06$ & $0.16 \pm 0.06$ & $+0.33 \pm 0.07$ & $-0.47 \pm 0.18$ \\
 Ca & $-0.011 \pm 0.006$ & $-0.14 \pm 0.03$ & $+0.146 \pm 0.013$ & $+0.06 \pm 0.05$ & $0.48 \pm 0.06$ & $+0.34 \pm 0.07$ & $+0.02 \pm 0.09$ \\
 Sc & $-0.238 \pm 0.009$ & $+0.20 \pm 0.04$ & $-0.150 \pm 0.012$ & $+0.22 \pm 0.05$ &                 &                 &                  \\
 Ti & $-0.129 \pm 0.006$ & $+0.08 \pm 0.03$ & $+0.002 \pm 0.012$ & $+0.14 \pm 0.05$ & $0.39 \pm 0.05$ & $+0.23 \pm 0.07$ & $-0.18 \pm 0.09$ \\
 V  & $-0.235 \pm 0.016$ & $+0.09 \pm 0.09$ & $-0.095 \pm 0.017$ & $+0.24 \pm 0.11$ & $0.42 \pm 0.09$ & $+0.13 \pm 0.10$ & $-0.25 \pm 0.14$ \\
 Cr & $-0.359 \pm 0.010$ & $+0.13 \pm 0.05$ & $-0.081 \pm 0.024$ & $+0.47 \pm 0.09$ & $0.99 \pm 0.18$ & $+0.00 \pm 0.07$ & $+0.00 \pm 0.10$ \\
 Mn & $-0.579 \pm 0.013$ & $+0.08 \pm 0.06$ & $-0.341 \pm 0.019$ & $+0.23 \pm 0.07$ & $0.81 \pm 0.13$ & $-0.22 \pm 0.07$ & $-0.31 \pm 0.10$  \\
 Fe & $-0.359 \pm 0.006$ & $+0.06 \pm 0.03$ & $-0.079 \pm 0.020$ & $+0.39 \pm 0.08$ & 1.00 & 0.00 & 0.00 \\
 Co & $-0.261 \pm 0.012$ & $+0.09 \pm 0.07$ & $-0.162 \pm 0.014$ & $+0.19 \pm 0.09$ & $0.28 \pm 0.06$ & $+0.10 \pm 0.10$ & $-0.45 \pm 0.14$ \\
Ni & $-0.370 \pm 0.007$ & $+0.08 \pm 0.04$ & $-0.236 \pm 0.012$ & $+0.17 \pm 0.05$ & $0.40 \pm 0.06$ & $-0.01 \pm 0.10$ & $-0.41 \pm 0.12$ \\
 Cu & $-0.554 \pm 0.022$ & $+0.38 \pm 0.11$ & $-0.579 \pm 0.036$ & $+0.47 \pm 0.20$ &                 &  & \\
 Zn & $-0.231 \pm 0.009$ & $+0.06 \pm 0.04$ & $-0.152 \pm 0.015$ & $+0.05 \pm 0.06$ & $0.22 \pm 0.06$ & $+0.13 \pm 0.10$ & $-0.53 \pm 0.15$ \\
\noalign{\smallskip}
\hline
\end{tabular}
\end{table*}

\subsection{Empirical supernovae yields}
\label{yields}
Adopting notations used by \citet{weinberg17} in their analytic one-zone GCE models for 
the abundance of an element X in the interstellar gas
due to prompt CC SNe and delayed Ia SNe enrichment, the change in \xmg\ relative to 
the early-time ($t_0$) CC plateau is
\begin{eqnarray} 
\Delta \xmg \, = {\rm log} \, \frac{Z_{\rm X}^{\rm CC}(t) \, + \, Z_{\rm X}^{\rm Ia}(t)}{Z_{\rm Mg}^{\rm CC}(t)}
\, - \, {\rm log} \, \frac{Z_{\rm X}^{\rm CC}(t_0)}{Z_{\rm Mg}^{\rm CC}(t_0)}.
\end{eqnarray}
Here $Z_{\rm X}^{\rm CC}(t)$ and $Z_{\rm X}^{\rm Ia}(t)$ are the abundances of element X 
produced by CC and Ia SNe, respectively, and it is assumed that only CC SNe contribute to 
Mg.

Assuming that the IMF integrated CC yield, $m_{\rm X}^{\rm CC}$,
is constant in time, that is, independent of metallicity, we have
\begin{eqnarray}
\frac{Z_{\rm X}^{\rm CC}(t)}{Z_{\rm Mg}^{\rm CC}(t)} \, = \,
\frac{Z_{\rm X}^{\rm CC}(t_0)}{Z_{\rm Mg}^{\rm CC}(t_0)} \, = \,
\frac{m_{\rm X}^{\rm CC}}{m_{\rm Mg}^{\rm CC}}.
\end{eqnarray}
Equation\,(1) then yields
\begin{eqnarray}
\Delta \xmg \, = {\rm log} \, \left(\,1 \, + \, \frac{Z_{\rm X}^{\rm Ia}(t)}{Z_{\rm X}^{\rm CC}(t)}\right),
\end{eqnarray}
and as a special case with X\,=\,Fe 
\begin{eqnarray}
\Delta \femg \, = {\rm log} \, \left(1 \, + \, \frac{Z_{\rm Fe}^{\rm Ia}(t)}{Z_{\rm Fe}^{\rm CC}(t)} \right),
\end{eqnarray}
which corresponds to Eq.\,(42) in \citet{weinberg17} when replacing Mg with O  as the
CC reference element.

Assuming that the SNe Ia yield, $m_{\rm X}^{\rm Ia}$, is also independent of metallicity
and introducing the ratio between Ia and CC SNe yields divided by the same ratio for Fe, i.e.
\begin{eqnarray}
F_{\rm X} \, = \, \frac{m_{\rm X}^{\rm Ia} / m_{\rm X}^{\rm CC}} 
                       {m_{\rm Fe}^{\rm Ia} / m_{\rm Fe}^{\rm CC}}, 
\end{eqnarray}
we get
\begin{eqnarray}
\Delta \xmg \, = {\rm log} \, \left(\,1 \, + \, F_{\rm X} (10^{\rm \Delta [Fe/Mg]} - 1)\, \right).
\end{eqnarray}
corresponding to Eq.\,(10) in \citet{feuillet18} when Mg is replaced with O.

Solving Eq.\,(6) with respect to $F_{\rm X}$ we get
\begin{eqnarray}
F_{\rm X} \, = \, \frac{10^{\rm \Delta [X/Mg]} - 1}{10^{\rm \Delta [Fe/Mg]} - 1}.
\end{eqnarray}

Introducing logarithmic yields, $\xfe _{\rm CC}$ and $\xfe _{\rm Ia}$,
we get from Eq. (5)
\begin{eqnarray}
\xfe _{\rm Ia} \, = \, \xfe _{\rm CC} \, + \, {\rm log} \, F_{\rm X}.
\end{eqnarray}

$\xfe _{\rm CC}$ is determined as the mean value of \xfe\ for the high-$\alpha$
stars and is given in Col.\,7 of Table \ref{table:fits}. For elements having a significant
slope of \xmg\ versus \mgh\ (Na, Sc, and Cu) no value is given. These elements
have metallicity dependent CC SNe yields and Eqs.\,(2) to (8) are therefore not valid.
Furthermore, the value of $\cfe_{\rm CC}$ is omitted, because carbon probably
gets a significant contribution from intermediate-mass AGB stars at $\feh \simgt -1.5$
\citep[][Fig. 5]{kobayashi20}.
   
The standard deviation of the mean value of $\xfe _{\rm CC}$ ranges from 0.005\,dex for Ni 
to 0.013\,dex for O, but systematic uncertainties are much larger. The 
high precision of the stellar abundances were obtained by  determining 
\teff , \logg , and \xfe\ spectroscopically relative to two bright standard stars,
HD\,22879 and HD\,76932. Hence, the error of $\xfe _{\rm CC}$ is determined
by the uncertainty of the standard star abundances relative to the solar abundances. Due to      
the uncertainty of the photometric parameters of the standard stars
($\sigma \teff = \pm 50$\,K and $\sigma \logg = \pm 0.05$\,dex) 
and the difficulty of measuring accurate equivalent widths of
lines in the crowded solar flux spectrum, 
we estimate that the error of $\xfe _{\rm CC}$ is  $\sim \! 0.05$\,dex.
In addition, there is a potential error of $\xfe _{\rm CC}$ for elements for which  
the derived abundances were not corrected for 3D non-LTE effects. In the case of elements 
having full 3D non-LTE corrections available, we find that such corrections lead to 
changes of +0.038\,dex for $\ofe _{\rm CC}$
and +0.019\,dex for  $\mgfe _{\rm CC}$. For elements with 1D non-LTE corrections,
the changes of $\xfe _{\rm CC}$ are +0.002\,dex for Si, +0.028\,dex for Ca, and as much as
+0.081\,dex for Mn. Guided by these numbers, we adopt errors of
$\xfe _{\rm CC}$ ranging from $\pm 0.06$\,dex for elements with full 3D non-LTE corrections
(O and Mg) and $\pm 0.10$\,dex for elements with no correction (V, Co, Ni, and Zn).  
For the remaining elements that have non-LTE corrections (Si, Ca, and Mn) or have
abundances determined from ionised lines (Ti and Cr), we adopt an error of $\pm 0.07$\,dex
for $\xfe _{\rm CC}$.

The $F_{\rm X}$ values are determined from Eq.\,(7) using ${\rm \Delta [X/Mg]}$ measured 
at $\mgh \! = \! -0.8$, that is, the difference of the $a$-coefficients for the linear
fits to the low-$\alpha$ and high-$\alpha$ stars. 
The resulting values are given in Col. 6 of Table \ref{table:fits}
together with the 1-$\sigma$ errors derived from the errors of the $a$-coefficients.  
In this connection, we note that the systematic errors of \xfe\ discussed above
do not affect ${\rm \Delta [X/Mg]}$ significantly, because low-$\alpha$ and high-$\alpha$ stars 
are affected in the same way.

Finally, empirical Ia SNe yields are determined from Eq.\,(8). Values are given in
Col. 8 of Table \ref{table:fits} except for O and Mg. Magnesium was assumed to be a pure
CC element, i.e. $F_{\rm X} = 0$, and the $F_{\rm X}$ value of O is not significantly 
different from zero indicating that the Ia yield is very small relative to the CC yield.

\begin{figure}
\centering
\resizebox{\hsize}{!}{\includegraphics{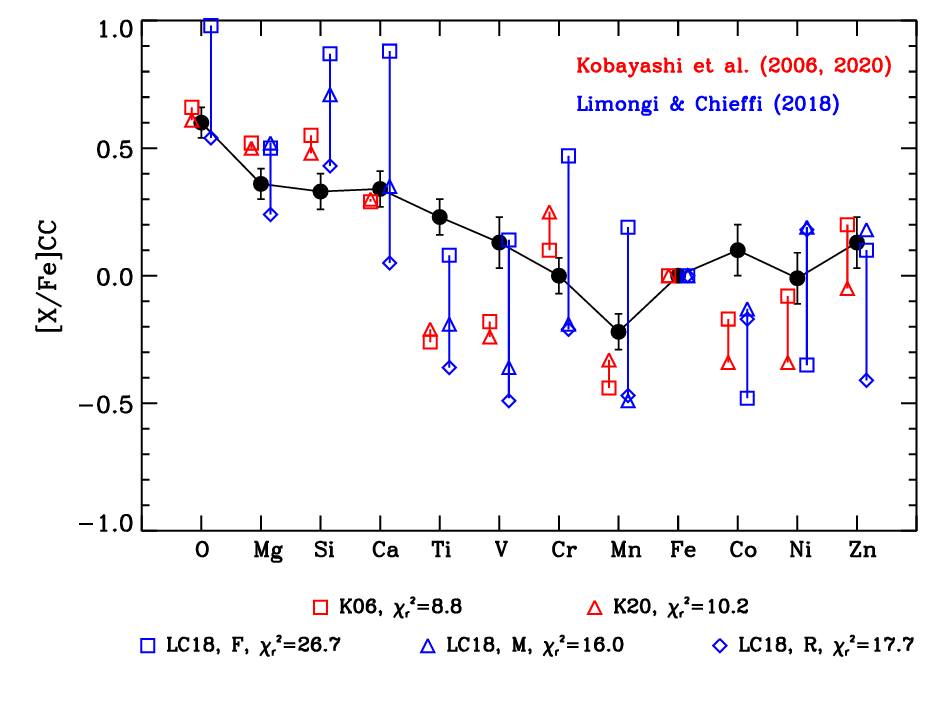}}
\caption{Comparison of empirical $\xfe _{\rm CC}$ yields with
IMF integrated yields predicted from
models of CC SNe. The empirical yields are shown with filled black circles.
Yields predicted from the models of \citet[][K06 and K20]{kobayashi06, kobayashi20}
are plotted with red symbols and yields calculated from
\citet[][LC18, models F, M, and R]{limongi18} with blue symbols; see text for details.
For each model, the reduced chi-square for the comparison with the empirical yields is given.}
\label{fig:yields.CC}
\end{figure}

\begin{figure}
\centering
\resizebox{\hsize}{!}{\includegraphics{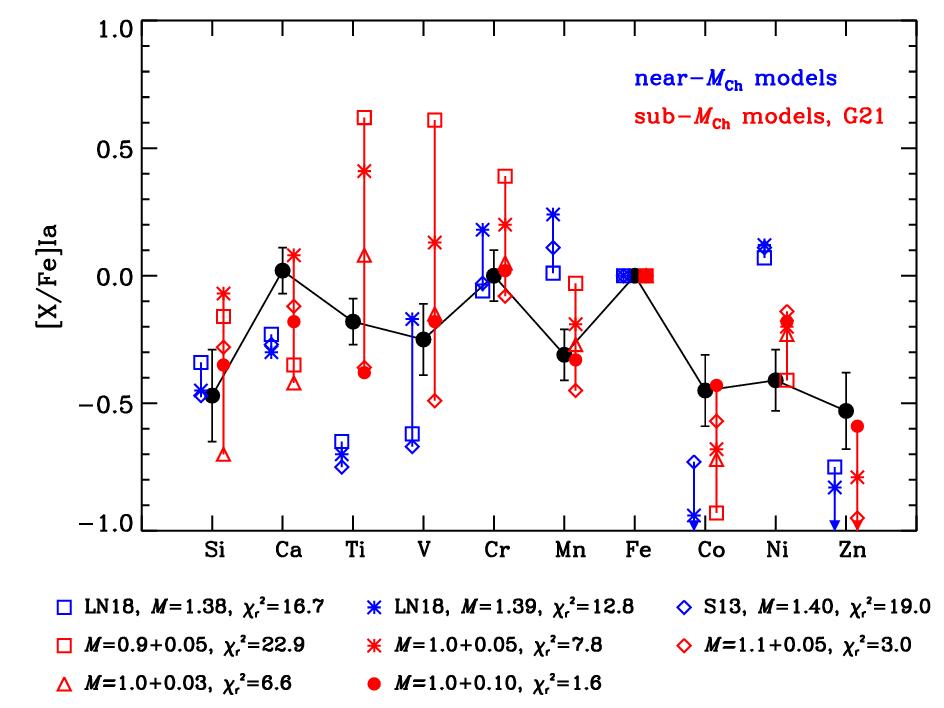}}
\caption{Comparison of empirical $\xfe _{\rm Ia}$ yields with yields predicted from
models of Ia SNe. The empirical yields are shown with filled black circles.
Yields predicted from near-\Ch\ models of \citet[][LN18]{leung18} and
\citet[][S13]{seitenzahl13} are plotted with blue symbols and yields calculated from
sub-\Ch\ models of  \citet[][G21]{gronow21} with red symbols.
The core masses and masses of the He envelope for the sub-\Ch\ models
are given in units of the mass of the Sun. For each model, the reduced chi-square for
comparison with the empirical yields is given.}
\label{fig:yields.Ia}
\end{figure}

\subsection{Comparison of empirical and predicted SNe yields}
\label{predicted}
A complete comparison between our empirical yields and yields calculated for models of SNe
is outside the scope of this paper. We may, however, provide some examples
to show how the empirical yields can be used to constrain parameters of the models.
As a measure of the success of the comparison we use the reduced chi-square 
\begin{eqnarray}
\chi_r^2 \,= \, \frac{1}{N} \, \sum_{i=1}^{N} \, (\xfe _{i,\rm pre} - \xfe _{i,\rm emp})^2\,/\,\sigma _{i, \rm emp}^2,
\end{eqnarray}
where $\xfe _{i,\rm pre}$ and $\xfe _{i,\rm emp}$ are the predicted and empirical yields, respectively,
for an element $i$, and
$\sigma _{i, \rm emp}$ is the estimated error of the empirical yield.
$N$ is the number of elements, i.e. $N\!=\!11$ for the CC yields and $N\!=\!9$ for the Ia
yields according to Table \ref{table:fits}.

The empirical CC yields were compared with yields calculated by \citet{kobayashi06, kobayashi20} 
and \citet{limongi18} for spherical models with a metallicity
of $Z = 0.1 Z_{\sun}$, close to the mean $Z$ for our stars. As discussed in these
papers, assumptions about formation of remnants (neutron stars or black holes) 
are critical for the yield calculations.  \citet{limongi18} have three sets of models:
`set F' in which the stars are assumed to eject $0.07 \, \Msun$ $^{56}$Ni
(decaying to $^{56}$Fe),
`set M', which is based on the mixing and fallback scheme of \citet{umeda02},
and `set R', which is the same as `set M' for stars with masses in the range
13\,-\,25\,\Msun, while more massive stars are assumed to collapse to black holes and only 
contribute to the yields via stellar winds. Total yields were obtained by 
integrating over the \citet{salpeter55} IMF from 13 to 120\,\Msun\ and are converted to 
\xfe\ values using the \citet{asplund21} solar abundances. In \citet[][K06]{kobayashi06}, the
mass cuts are chosen so that 0.07\,\Msun\ $^{56}$Fe is produced by normal Type II SNe,
and yields from high-energy SNe, so-called hypernovae (HNe), are included. 
The parameters of the
mixing and fallback scheme for HNe are constrained to provide a yield of  \ofe\,=\,0.5
and a relative HNe fraction of 0.5 for $M \ge 20\,\Msun$ is assumed. Total yields are found 
by integration over the Salpeter IMF from 13 to 50\,\Msun . The yields in \citet[][K20]{kobayashi20} represent
an update of the K06 yields and adoption of the \citet{kroupa08} IMF.

As seen from Fig. \ref{fig:yields.CC}, all models give a poor fit to the empirical
CC yields, i.e. $\chi_r^2 \gg 1$. The Limongi \& Chieffi models show large differences
in the yields\,\footnote{We have adopted yields 
from \citet{limongi18} for models with an initial rotation velocity $V = 0$\,\kmprs ; models
with higher values, $V = 150$ and 300\,\kmprs , have similar high $\chi_r^2$ values.} 
and have $\chi_r^2 \ge 16$. The Kobayashi et al.
models perform somewhat better, but the predicted yields of Ti, V, and Co are much lower
than the empirical yields leading to
$\chi_r^2 \sim 10$. As discussed by \citet{kobayashi20}, the yields of Ti, V, and Co are 
sensitive to multidimensional effects. Based on 2D models with jets 
\citep{maeda03, tominaga09}, it was estimated that the yields of Ti, V, and Co 
should be increased by 0.45, 0.2, and 0.2\,dex, respectively. This 
decreases the reduced chi-square to $\chi_r^2 \sim 4$.

Figure \ref{fig:yields.Ia} shows a comparison of the empirical Ia SNe yields
with yields calculated for near-\Ch\ and sub-\Ch\ models having  $Z = 0.1 \, Z_{\sun}$. 
Included are yields from \citet{leung18} for two near-\Ch\ deflagration-to-detonation models 
with masses at explosion,
$M$\,=\,1.38 and $1.39\,\Msun$, and hence different core densities
($\rho _{\rm c} = 3\,\times \, 10^9$\,g\,cm$^{-3}$ and $\rho _{\rm c} = 5\, \times \,10^9$\,g\,cm$^{-3}$), 
and yields from a similar model 
with  $M\,=\,1.40 \,\Msun$  and $\rho _{\rm c} = 2.9 \times \, 10^9$\,g\,cm$^{-3}$ by
\citet{seitenzahl13}. The sub-\Ch\ yields all refer to models by
\citet{gronow21} of double detonation white dwarfs with different masses of the C-O core
and the He-shell as shown in the figure. 

As seen from Fig. \ref{fig:yields.Ia}, the yields of the three near-\Ch\ models give
a poor fit to the empirical yields; the predicted yields for Ca and Ti are too low and
those of Mn and Ni are too high, leading to $\chi_r^2 > 12$. The sub-\Ch\ models
of \citet{gronow21} have large differences in the predicted yields, depending on the masses of
the C-O core and the He-shell, but  
the model with a core mass of $M \! = \!1.0 \,\Msun$ and a He-shell mass of $M\!=\!0.10\,\Msun$
has a  reduced chi-square of $\chi_r^2 = 1.6$ only. As seen, the  
predicted yields for this model (filled red circles in Fig. \ref{fig:yields.Ia}) 
do not deviate more than two-sigma from the
empirical yields. This suggests that a dominant contribution 
from sub-\Ch\ Ia SNe is needed to explain the chemical evolution of the GSE dwarf
galaxy as was also found by \citet{sanders21} based on \mnfe\ and \nife\ ratios.

\subsection{Comparison to dSph galaxies}
\label{dSph}
The abundance pattern of the low-$\alpha$ population, which comes from disrupted dwarf galaxies,
is in good general agreement with results from surviving satellite dwarf spheroidal 
(dSph) galaxies around the Milky Way; at least for those dSph galaxies with relatively
short star formation histories, comparable to that expected for GSE.
A detailed abundance pattern for stars in the Sculptor dSph galaxy was provided in the work
of \citet{Hill2019}. They used the increase in [X/Mg] with [Fe/H] to estimate the relative Ia SNe
contribution to each element (their Fig.~14). They found the strongest impact of Ia SNe on the
elements Cr, Mn, Fe, which showed similar slopes in excellent agreement
with our results (Fig.\,\ref{fig:xmg-mgh}). Furthermore, they found a modest but significant
contribution of Ia SNe to the elements S, Ca, Ti, Sc, and Co, while the results for Ni and Zn
hinted at a small contribution but were less conclusive. This is consistent with the
low-$\alpha$ population presented here.

Analyses of dSph galaxies have also been used to study the nature of
low-metallicity Ia SNe.
\citet{Kirby2019} used the abundances of Mg, Si, Ca, Cr, Co, Ni, and Fe to try and
understand the impact of Ia SNe in five dSph galaxies: Sculptor, Leo~II, Draco, Sextans, and Ursa Minor.
Based mainly on the [Ni/Fe] abundances, \citet{Kirby2019} concluded that the dominant Ia SNe channel
in ancient dSph galaxies is that of sub-$M_{ch}$ white dwarfs, as found here via the low-$\alpha$
population; however this might differ for 
galaxies with more extended star formation histories. Other works have focused on fewer stars
and/or fewer elements, but those targeting Mn all suggested a significant contribution of Ia SNe to Mn,
pointing to sub-$M_{ch}$ white dwarfs as the dominant Ia SNe channel, with possibly also
significant contribution of Type Iax SNe \citep{North2012,Cescutti2017,delosReyes2020}.

\subsection{Disentangling PISN and Ia SNe}
\label{PISN}
The results presented in Fig.\,\ref{fig:xmg-mgh}  can be used to check for descendants of so-called
Pair Instability Supernovae (PISNe), which have been predicted to be the death of massive
zero-metallicity stars of $150 \le M_\star /\Msun \le 260$ \citep[e.g.][]{Heger2002}. 
The abundance pattern of such stars is very unique, with very little production of odd-$Z$ elements, 
resulting in a strong odd-even effect. The surviving descendants are predicted to be very rare, 
only $<0.1\%$ of Milky Way halo stars at $\rm[Fe/H]\approx-1$ are expected to have received 
$>50\%$ of their metals from PISNe \citep{deBennassuti2017}. Therefore it is unsurprising 
that no star in our sample shows evidence of such imprint.

Our sample can, however, be used as an empirical benchmark for the
abundance pattern of early Ia SNe,
which overlaps to a certain degree with that predicted for PISNe. In particular, low abundances
of the elements Na and Cu relative to Ca or Fe have been identified as a clear signature of PISNe
\citep{Salvadori2019}. However, our results show very clearly that
low-metallicity Ia SNe do not produce Na and Cu; the \xmg\ versus \mgh\ trends for these two elements
are the same for low- and high-$\alpha$ stars.
It is therefore clear that a star which would be dominantly enriched by an early Type Ia supernova,
without much CC SNe contribution, would also have low Na and Cu abundances.

There are, however, some key differences between the abundance patterns of PISNe and Ia SNe. 
As seen from Fig. 3 in \citet{Salvadori2019}, the production of Co and Zn by PISNe
in the mass range $150 \le M_\star /\Msun \le 260$
is very low compared to the production of Ni, i.e. $\coni _{\rm PISN} < -0.6$ 
and $\znni _{\rm PISN} < -1.3$.
As seen from Figs. \ref{fig:yields.CC} and \ref{fig:yields.Ia}, the empirical yields of
these abundance ratios are significantly higher both for CC and Ia SNe, i.e.
$\coni _{\rm CC} > -0.10$, $\coni _{\rm Ia}  > -0.30$,
$\znni _{\rm CC} > -0.10$, and $\znni _{\rm Ia}  > -0.40$.
Hence, it should be possible to find signatures
of PISNe by measuring accurate Co, Ni, and Zn abundances. We note, in this connection, that
two stars, for which imprints of PISNe have been suggested, i.e.
SDSS J0018-0939 \citep{aoki14} and LAMOST 1010+2358 \citep{Xing2023}, have
$\coni = -0.77 \pm 0.15$ and $\coni = -0.55 \pm 0.08$. 
Zn abundances of the two stars were not determined, only upper limits,
and Co and Ni abundances were based on a 1D LTE analysis.
Clearly, 3D non-LTE abundances of Co, Ni, and Zn for
the PISNe imprint candidates and the high-$\alpha$ and low-$\alpha$ stars used to determine CC
and Ia yields would be of high interest.

\section{Summary and conclusions}
\label{Summary}
In this paper, we have extended previous determinations \citep{nissen10, nissen11}
of 1D LTE elemental abundances in high-$\alpha$ and low-$\alpha$ halo stars to include Sc, V, and Co,
so that the nucleosynthesis of all iron-peak elements from Sc to Zn can be studied 
along with the lighter elements C, O, Na, Mg, Si, and Ca. 

In order to improve the accuracy of the abundance determinations, new 3D non-LTE corrections
for Mg and 1D non-LTE corrections for Na, Si, Ca, Mn, and Cu were applied. 
For C and O, 3D non-LTE abundances from \citet{amarsi19b} were used,
and for Ti, Cr, and Fe, we used spectral lines from the ionised species, for which non-LTE corrections of
the derived abundances are negligible. 

After having applied the 3D non-LTE corrections for the Mg abundances, substructure in the
\femg\,-\,\mgh\ diagram of the low-$\alpha$ halo stars emerged (see Fig. \ref{fig:fe2-all}). 
A group of six stars 
with $\mgh \sim -0.8$ have higher values of \femg\ than a group of seven stars 
with similar metallicity. A similar splitting is seen for \crmg\ and \fena . Furthermore,
there is a striking difference in the kinematics of the two groups suggesting that stars in the 
high-\femg\ group have been accreted from the GSE dwarf galaxy and stars in the 
low-\femg\ group from the Thamnos galaxy.   

The \xmg\ versus \mgh\ trends for high-$\alpha$ and low-$\alpha$ halo stars (Fig. \ref{fig:xmg-mgh})
were used to study the nucleosynthesis of the elements.
By fitting straight lines to the two populations (excluding Thamnos stars from
the low-$\alpha$ population), we determined empirical $\xfe _{\rm CC}$ yields from
the high-$\alpha$ trends and then $\xfe _{\rm Ia}$ yields from the difference in the trends
of the low-$\alpha$ and high-$\alpha$ populations assuming that Mg is a pure CC element. 
C, O, Na, and Cu have negligible contributions from Ia SNe, whereas  
Cr, Mn, and Fe have the largest contribution (see the $F _{\rm X}$ values in
Table \ref{table:fits}). This agrees with recent results for surviving
dSph galaxies \citep{Hill2019, Kirby2019}.     

The estimated uncertainties of the empirical yields range from $\pm 0.06$ to $\pm 0.18$\,dex
of which the main part arises from possible systematic errors in the stellar abundances relative to 
the solar abundances and from the lack of 3D and non-LTE corrections for some of the
elements. For elements having such corrections, the effect on the derived \xfe\ yields 
ranges from 0.00\,dex for Si to +0.08\,dex for Mn. We conclude that it would be important to
perform 3D non-LTE calculations for other elements (Sc, V, Co, Ni, and Zn in particular) and also to update
the corrections for Mn based on full 3D non-LTE calculations.

Predictions of yields from spherical
models of CC SNe by \citet{kobayashi06, kobayashi20} and \citet{limongi18}
do not agree well with the empirical $\xfe _{\rm CC}$ yields; the reduced chi-square
of the comparison is $\chi _r^2 > 8$ (see Fig. \ref{fig:yields.CC}). Hopefully,       
the empirical yields can help to constrain
free parameters of such models, for example the mass limit for formation of remnants
(black wholes or neutron stars). As discussed by \citet{kobayashi20}, it may, however,
be necessary to consider
multidimensional models, for example the 2D models with jets by \citet{maeda03} and \citet{tominaga09},
to explain the empirical CC yields of Ti, V, and Co.  

The predicted yields for near-\Ch\ models of Ia SNe by \citet{leung18} and
\citet{seitenzahl13} give a poor fit to the empirical $\xfe _{\rm Ia}$ values, i.e.
$\chi _r^2 > 12$ (see Fig. \ref{fig:yields.Ia}). These models predict too high \mnfe\ and
\nife\ as well as too low \cafe\ and \tife\ values. Double detonation sub-\Ch\ models
by \citet{gronow21} prowide, however, an excellent fit ($\chi _r^2 \sim 1.6$) for the
right choice of the masses of the C-O core and the He shell. This suggests that sub-\Ch\
Ia SNe play a dominant role in the chemical evolution of the GSE dwarf galaxy.
The same conclusion was reached by \citet{sanders21} based on Mn and Ni abundances in 
the GSE stars. Our conclusion is based on empirical \xfe\ yields for nine elements of which
Ca, Ti, Mn, and Ni are the most important in favour of sub-\Ch\ models.

Finally, we note that precise knowledge of yields of CC and Ia SNe is important 
when searching for signatures of massive Pair Instability Supernovae.
In particular, it seems that low \coni\ and \znni\ ratios can be used
to identify imprint of PISNe on the abundances of metal-poor stars.

\begin{acknowledgements}
We thank the referee for important comments and
Chiaki Kobayashi for providing integrated yields for the CC models 
in \citet{kobayashi20}.
A.M.A. acknowledges support from the Swedish Research Council (VR 2020-03940). 
\'{A}.S. has received funding from the European Research Council (ERC) under 
the European Union’s Horizon 2020 research and innovation programme (grant agreement No 804240).
W.J.S. wishes to thank Edilberto Sanchez Moreno of the Computing Center for considerable assistance.
This research was supported by computational resources
provided by the Australian Government through the National Computational Infrastructure
(NCI) under the National Computational Merit Allocation Scheme and the ANU Merit
Allocation Scheme (project y89).
This research has made use of data from the European Space Agency (ESA) mission
{\it Gaia} (\url{https://www.cosmos.esa.int/gaia}), processed by the {\it Gaia}
Data Processing and Analysis Consortium (DPAC,
\url{https://www.cosmos.esa.int/web/gaia/dpac/consortium}). Funding for the DPAC
has been provided by national institutions, in particular the institutions
participating in the {\it Gaia} Multilateral Agreement.
This research has made use of the SIMBAD database operated at CDS, Strasbourg, 
France \citep{wenger00}.
\end{acknowledgements}

\bibliographystyle{aa}
\bibliography{nissen.2023}

\end{document}